\pdfoutput=1
\documentclass[prl,aps,twocolumn,superscriptaddress,showpacs]{revtex4-1}

\usepackage{graphicx}% Include figure files
\usepackage{epstopdf}
\usepackage{amsmath}

\begin{document}

\title{Magnetism of the Fe$^{2+}$ and Ce$^{3+}$ sublattices in Ce$_{2}$O$_{2}$FeSe$_{2}$: a combined neutron powder diffraction, inelastic neutron scattering and density functional study}

\author{E. E. McCabe}
\affiliation{School of Physical Sciences, University of Kent, Canterbury, CT2 7NH, UK}
\affiliation{Department of Chemistry, Durham University, Durham, DH1 3LE, UK}
\author{C. Stock}
\affiliation{School of Physics and Astronomy, University of Edinburgh, Edinburgh EH9 3JZ, UK}
\author{J. L. Bettis Jr.}
\affiliation{Department of Chemistry, North Carolina State University, Raleigh, North Carolina, 27695-8204, USA}
\author{M.-H. Whangbo}
\affiliation{Department of Chemistry, North Carolina State University, Raleigh, North Carolina, 27695-8204, USA}
\author{J. S. O. Evans}
\affiliation{Department of Chemistry, Durham University, Durham, DH1 3LE, UK}
\date{\today}

\begin{abstract}

The discovery of superconductivity in the 122 iron selenide materials above 30 K necessitates an understanding of the underlying magnetic interactions.  We present a combined experimental and theoretical investigation of magnetic and semiconducting Ce$_{2}$O$_{2}$FeSe$_{2}$ composed of chains of edge-linked iron selenide tetrahedra. The combined neutron diffraction and inelastic scattering study and density functional calculations confirm the ferromagnetic nature of nearest-neighbour Fe -- Se -- Fe interactions in the ZrCuSiAs-related iron oxyselenide Ce$_{2}$O$_{2}$FeSe$_{2}$. Inelastic measurements provide an estimate of the strength of nearest-neighbor Fe -- Fe and Fe -- Ce interactions. These are consistent with density functional theory calculations, which reveal that correlations in the Fe--Se sheets of Ce$_{2}$O$_{2}$FeSe$_{2}$ are weak. The Fe on-site repulsion $U_{Fe}$ is comparable to that reported for oxyarsenides and K$_{1-x}$Fe$_{2-y}$Se$_{2}$, which are parents to iron-based superconductors.

\end{abstract}

\maketitle

\section{1. Introduction}

The discovery of iron-based superconductivity \cite{Kamihara-2008,Paglione-2010,Stewart-2011,Johnston-2010} with transition temperatures as high as 55 K \cite{Ren-2008} has prompted efforts to understand both the electronic structure and magnetism of these materials, which are interrelated with superconductivity \cite{Wysocki-2011,Birgeneau-2006,Kastner-1998}. The first class of iron-based superconductors reported, the 1111 family, derive from $Ln$FeAsO ($Ln$ = trivalent lanthanide). They adopt the ZrCuSiAs structure \cite{Johnson-1974}, composed of layers of edge-sharing O$Ln_{4}$ tetrahedra alternating with layers of edge-sharing FeAs$_{4}$ tetrahedra. A second class, 122 materials, derive from $A$Fe$_{2}$As$_{2}$ ($A$ = Ca, Ba) with the ThCr$_{2}$Si$_{2}$ structure \cite{Ban-1965}, which again contains layers of edge-sharing FeAs$_{4}$ tetrahedra. The metallic parent phases in both classes undergo structural phase transitions from tetragonal to orthorhombic symmetry just above an antiferromagnetic (AFM) ordering temperature ($T_{\mathrm{N}}$ = 137 K for LaOFeAs \cite {Cruz-2008} and ~172 K for CaFe$_{2}$As$_{2}$ \cite{Goldman-2008}) with small ordered moments on the Fe sites in the $ab$ plane. Superconductivity has also been observed in the binary iron chalcogenide systems: the properties of Fe$_{1+x}$Te are very sensitive to the iron content \cite{Rodriguez-2010} and superconductivity can be induced by S or Se doping \cite{Mizuguchi-2009}; $\alpha$-FeSe does not order magnetically and undergoes a transition to a superconducting state at 8 K at ambient pressure \cite{Hsu-2008}, or 37 K at 7 GPa \cite{Margadonna-2009}. Recently, attention has turned to the potassium-iron-selenide phase diagram, in particular, K$_{0.8}$Fe$_{1.6}$Se$_{2}$, which adopts a vacancy-ordered ThCr$_{2}$Si$_{2}$ structure. This material is semiconducting \cite{Guo-2010,Yan-2011,Fang-2012,Lei-2011,Zhao-2012,Petrovic-2011,Petrovic-2011b} and orders antiferromagnetically below 559 K. Interestingly, the ordered Fe$^{2+}$ moments are large (3.31  $\mu_{B}$) and are oriented perpendicular to the layer, in contrast to the 1111 and 122 materials \cite{Bao-2011}. 

The magnetism of the iron sublattice in these materials has been the focus of much study in recent years. Initial studies on the 1111 and 122 materials suggested that the observed stripe magnetic ordering (ferromagnetic stripes along [010] in the orthorhombic unit cell) \cite{Zhao-2008,Fang-2008} arises from the competing nearest-neighbor (nn) and next-nearest-neighbor (nnn) AFM interactions \cite{Yildirim-2008,Si-2008}. Subsequent work has highlighted the roles of other factors which lead to the complexity of the magnetic phase diagram for these materials \cite{Wysocki-2011,Li-2012}.

In $Ln$FeAsO and $Ln$MnAsO materials, the $Ln^{3+}$ ions have a significant role not only in tuning the superconducting transition temperature in the doped phases (e.g. $T_{\mathrm{c}}$ = 26 K for LaFeAsO$_{1-x}$F$_{x}$ \cite{Kamihara-2008}, and 55 K for SmFeAsO$_{1-x}$F$_{x}$ \cite{Ren-2008}), but also in influencing the magnetism in the undoped parent phases. For example, the Fe$^{2+}$ moments of CeFeAsO order antiferromagnetically in the $ab$ plane at $T_{\mathrm{N,Fe}}$ = 140 K \cite{Zhao-2008} while the Ce$^{3+}$ moments couple strongly with the Fe$^{2+}$ moments at relatively high temperatures \cite{Maeter-2009}, before developing a long range order below $\sim$3.7 K with moments predominantly in the $ab$ plane \cite{Zhao-2008}. Recent studies suggested some reorientation of the Fe moments within the $ab$ plane at the onset of the long range order of the Ce moments \cite{Zhang-2013}. The Ce$^{3+}$ ions  influence the iron magnetic sublattice, and can also induce exotic properties such as Kondo screening of the local moment in closely-related CeFePO \cite{Bruning-2008} and CeRuPO \cite{Krellner-2007}.

\begin{figure}[t] 
\includegraphics[width=1.0\linewidth,angle=0.0]{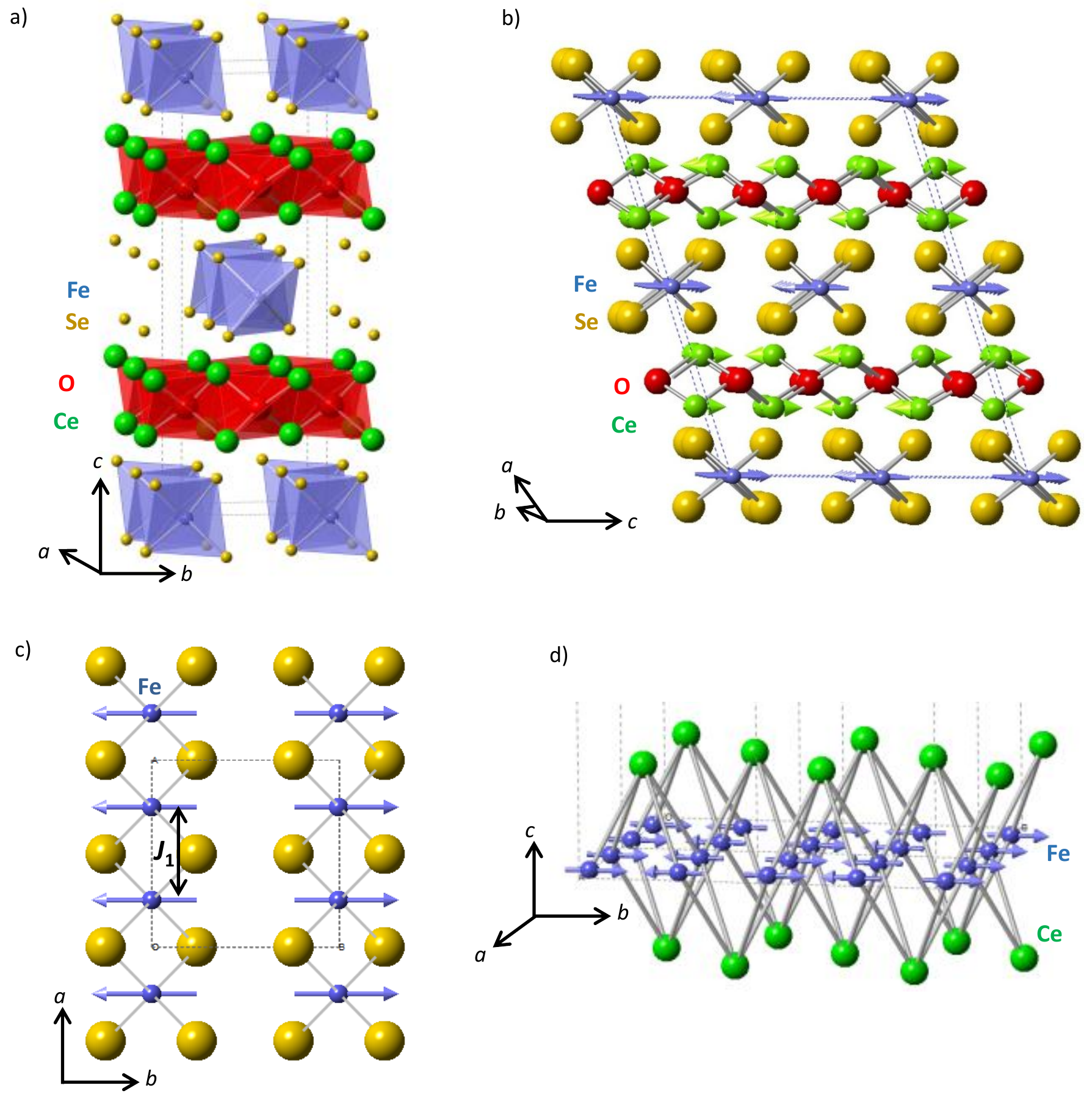}
\caption{[color online]  (a) Orthorhombic nuclear unit cell and (b) monoclinic magnetic unit cell of Ce$_{2}$O$_{2}$FeSe$_{2}$ (Ce = green, Fe = blue, O = red, and Se = yellow spheres). (c) An isolated sheet of edge-sharing chains of FeSe$_{4}$ tetrahedra present in Ce$_{2}$O$_{2}$FeSe$_{2}$. (d) Zoomed-in view of the magnetic ordering in the Fe and Ce sublattices of Ce$_{2}$O$_{2}$FeSe$_{2}$. For convenience of discussion, the directions of the orthorhombic unit cell are used to describe the magnetic structure in (c) and (d); the FM chains of edge-sharing FeSe$_{4}$ tetrahedra lie in the $ab$-plane with the FM chains running along the $a$-direction.}
\label{structure}
\end{figure}

The synthesis and crystal/magnetic structures of the iron oxyselenide Ce$_{2}$O$_{2}$FeSe$_{2}$ were reported in 2011 \cite{McCabe-2011}. It adopts a ZrCuSiAs-related structurein which the transition metal sites are half occupied by Fe$^{2+}$ cations in a stripe ordered structure (Fig.\ref{structure}$a$). The magnetic structure of Ce$_{2}$O$_{2}$FeSe$_{2}$ (Fig.\ref{structure}$b$) determined from neutron powder diffraction (NPD) data reported rather surprising observations \cite{McCabe-2011}. It undergoes an AFM ordering below $T_{\mathrm{N}}$ = 171 K in which the Fe$^{2+}$ spins have a ferromagnetic (FM) order within each chain of edge-sharing FeSe$_{4}$ tetrahedra despite the Fe-Se-Fe angle (71.94$^{\circ}$) deviating strongly from 90$^{\circ}$ (Fig.\ref{structure}$b$), so one would have expected an AFM ordering according to the Goodenough-Kanamori rule \cite{Goodenough-1963,Goodenough-1955,Kanamori-1959}. In the present work we re-examine the magnetic ordering in Ce$_{2}$O$_{2}$FeSe$_{2}$ to confirm these unusual observations on the basis of NPD and inelastic neutron scattering (INS) experiments as well as density functional theory (DFT) calculations.  The paper is divided into five sections including this introduction; experimental and calculation descriptions; experimental and theoretical results;and finally a discussion and conclusion.

\section{2. Experimental details}
Ce$_{2}$O$_{2}$FeSe$_{2}$ was prepared as a black, polycrystalline sample (2.48 g) as described previously \cite{McCabe-2011}. Preliminary characterisation was carried out using a Bruker D8 X-ray diffractometer (reflection mode, Cu K$\alpha_{1}$/K$\alpha_{2}$ radiation, Lynxeye Si strip position sensitive detector, step size 0.02$^{\circ}$ with variable slits) equipped with an Oxford Cryosystems PheniX cryostat. NPD data were collected on the high-flux D20 diffractometer at Institut Laue Langevin (Grenoble, France) with neutron wavelength 2.41 \AA. The sample was placed in a 6 mm cylindrical vanadium can (to a height of $\sim$4 cm) and cooled to 2 K. Data were collected over a 2$\theta$ range of 5-130$^{\circ}$ at 2 K intervals on warming to 200 K. Powder diffraction data were analyzed by the Rietveld method \cite{Rietveld-1969} using the TOPAS Academic software suite \cite{Coelho-2003,Coelho-2012} controlled by local routines. The diffractometer zero point and neutron wavelength were initially refined using data collected at 12 K with lattice parameters fixed at values determined previously \cite{McCabe-2011}. The zero point and wavelength were then fixed in all subsequent refinements. Typically, the background was refined for each data set as well as the unit cell parameters and a Caglioti description of the peak shape. Structural characterization using data collected on the HRPD diffractometer at ISIS revealed no structural changes in this temperature range (4 -- 218 K) \cite{McCabe-2011}, so the atomic coordinates were fixed and this work focuses on the magnetic ordering. The web-based ISODISTORT software \cite{Campbell-2006} was used to obtain a magnetic symmetry mode description of the magnetic structure; magnetic symmetry mode amplitudes were then refined to determine the magnetic structures.

The same polycrystalline sample was used for INS measurements. The sample was packed into an Al foil envelope and placed in an Al can. Two experiments were performed using the MARI direct geometry chopper instrument at ISIS. The sample was cooled to 5 K in a closed-cycle cryostat. The energy of the incident beam, $E_{i}$, was selected using a Gd Fermi chopper spinning at 150 Hz (for $E_{i}$ = 40 meV) or 400 Hz (for $E_{i}$ = 150 meV).  In addition, a $t_{0}$ chopper was used to block fast neutrons and a thick disk chopper (spinning at 50 Hz) was used to improve background from neutrons above the Gd absorption edge. The cold triple-axis spectrometer SPINS at NIST Center for Neutron Research (Gaithersburg, USA) was used to investigate the temperature dependence of the crystal fields. A pyrolytic graphite (PG) monochromator (004 reflection) was used on the incident beam to give good resolution at high energy transfers and a PG(002) analyzer (horizontally focused over 11$^{\circ}$) was tuned to select a fixed final energy of E$_{f}$ = 5.0 meV. A Be filter was used on the scattered side. 

In our DFT electronic structure calculations for Ce$_{2}$O$_{2}$FeSe$_{2}$, we employed the projected augmented-wave (PAW) method encoded in the Vienna ab initio simulation package \cite{Kresse-1993,Kresse-1996,Kresse-1996b}, and the generalized gradient approximation (GGA) of Perdew, Burke and Ernzerhof \cite{Perdew-1996} for the exchange-correlation corrections, the plane wave cutoff energy of 500 eV, and the threshold of self-consistent-field (SCF) energy convergence of 10$^{-6}$ eV. We extract four spin exchange parameters by employing five ordered spin states defined on a ($a$, 2$b$, $c$) supercell (see below). The irreducible Brillouin zone was sampled with 4 $\times$ 2 $\times$ 1 $k$-points. To describe the electron correlation associated with the 3d states of Fe and the 4f states of Ce, the DFT plus on-site repulsion U (DFT+U) \cite{Dudarev-1998} calculations were carried out with effective $U_{eff}$ = $U$-$J$ (see below).

\section{3. Results}

In this section we outline the experimental and computational results of this paper.  We first discuss the neutron diffraction results probing the magnetic structure followed by a section discussing inelastic neutron results from which exchange constants between the Fe ions and the Ce ions are derived.  Finally, these are compared with density functional calculations.

\subsection{A. Neutron powder diffraction}

\begin{figure}[t] 
\includegraphics[width=1\linewidth,angle=0.0]{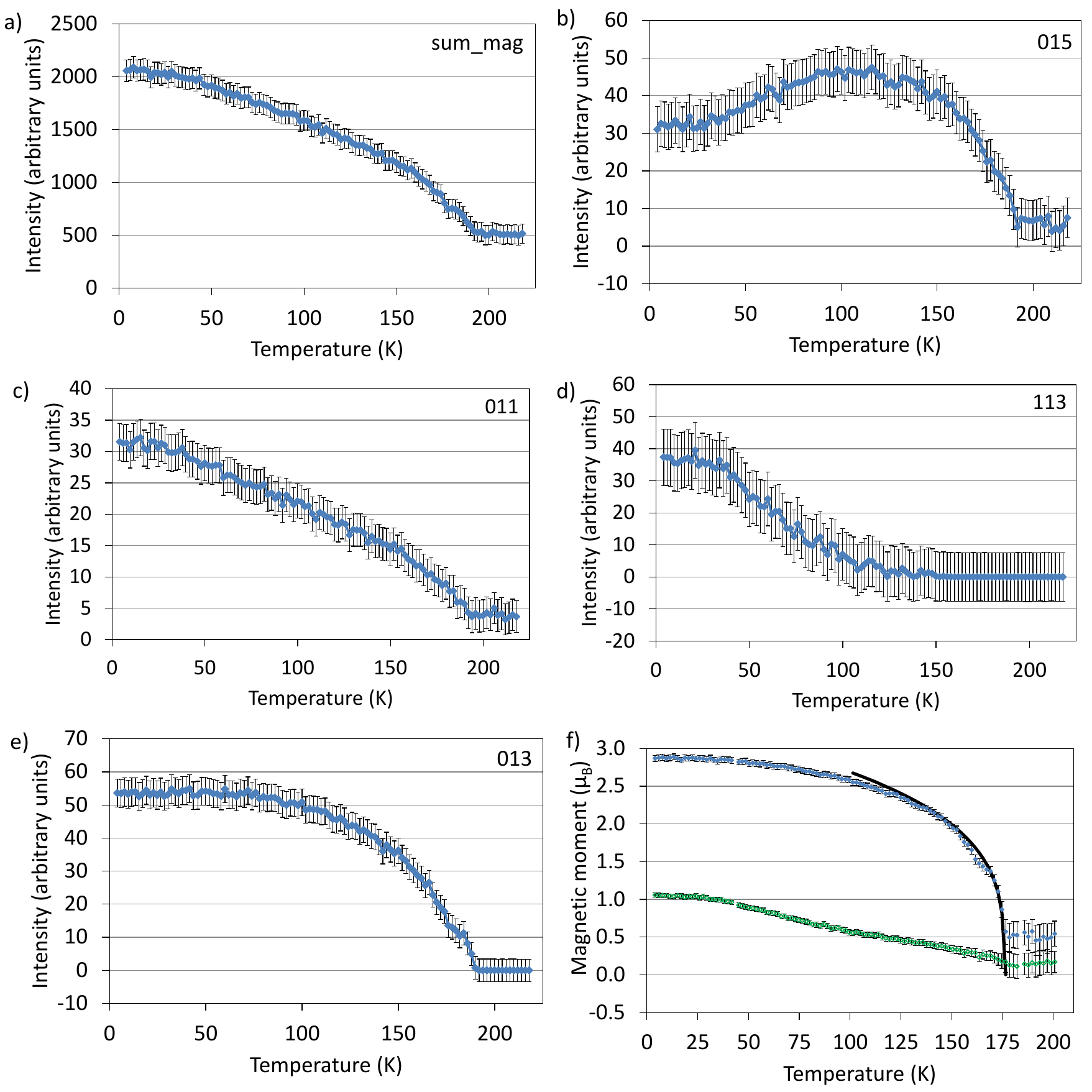}
\caption{[color online]  Temperature-dependence of the magnetic Bragg reflections: (a) Sum of the intensity of the 17 strongest magnetic reflections. (b-e) The intensity of the (0 1 1), (0 1 3), (0 1 5) and (1 1 3) reflections ($hkl$ indices refer to the $a_{n}$ $\times$ 2$b_{n}$ $\times$ 2$c_{n}$ magnetic unit cell) from the sequential refinements using a Pawley phase to fit the magnetic reflections. (f) Evolution of the Fe$^{2+}$ (blue, solid) and Ce$^{3+}$ (green, open) magnetic moments on cooling from Rietveld refinements. The solid black line shows fit to the function $M_{T}$ = $M_{0} (1-({T\over T_{N}}))^{\beta}$ for the Fe data between 100 - 171 K with $M_{0,Fe}$ = 3.40(4) $\mu_{B}$, $T_{N}$ = 175.5(8) K and $\beta$ = 0.28(1).}
\label{peaks}
\end{figure}

Rietveld analysis of NPD data collected at 250 K are consistent with the Fe-ordered, orthorhombic crystal structure described above. Additional reflections observed below $T_{N}$ are consistent with the magnetic ordering (and propagation vector $\vec{k}$ = (0 ${1\over 2}$ ${1\over 2}$)) reported previously \cite{McCabe-2011} and were indexed using an $a_{n}$ $\times$ 2$b_{n}$ $\times$ 2$c_{n}$ supercell (where the subscript $n$ refers to the nuclear unit cell); $hkl$ indices given subsequently for magnetic reflections refer to this magnetic unit cell. The intensity of these reflections increases smoothly on cooling to $\sim$100 K. Below this temperature, some reflections (e.g., (0 1 1), (0 1 9), (0 3 3)) continue to increase in intensity, others (e.g., (0 1 5), (0 1 7), (2 1 1)) decrease slightly (Fig.\ref{peaks}), while some additional very weak reflections (e.g. (1 1 1), (1 1 3)) are observed below this temperature.

The NPD data collected below $\sim$170 K can be fitted by the nuclear structure and a magnetic phase composed of FM chains of edge-sharing FeSe$_{4}$ tetrahedra, with AFM coupling between adjacent FM chains (Fig.\ref{structure}$b$). Attempts to fit the data with models containing AFM chains were not successful. The ``symmetry adapted ordering mode" approach \cite{Campbell-2006} was used here to describe the magnetically ordered structure. Mode inclusion analysis (described elsewhere, \cite{Tuxworth-2013}) was used to confirm that this arrangement of Fe moments gives the best fit to the data and does not change below $T_{N}$. Other models, including those with AFM chains, gave significantly worse fits. Whilst this FM-chain model gives magnetic Bragg peaks in the observed positions, the fit to the peak intensities was not perfect. Mode inclusion analyses were carried out at lower temperatures (80 K, 4 K) and confirmed that the arrangement of Fe$^{2+}$ moments does not change on cooling. Given the large ordered moment on the Fe sites in Ce$_{2}$O$_{2}$FeSe$_{2}$, we would expect our refinements to be sensitive to slight reorientations of the Fe$^{2+}$ moments, but there is no indication that reorientation of the Fe$^{2+}$ moments occurs. This is in contrast to the related PrFeAsO in which the Fe moments cant slightly along $c$ at the onset of Pr$^{3+}$ ordering \cite{Bhoi-2011}.

\begin{figure*}[t] 
\includegraphics[width=1.0\linewidth,angle=0.0]{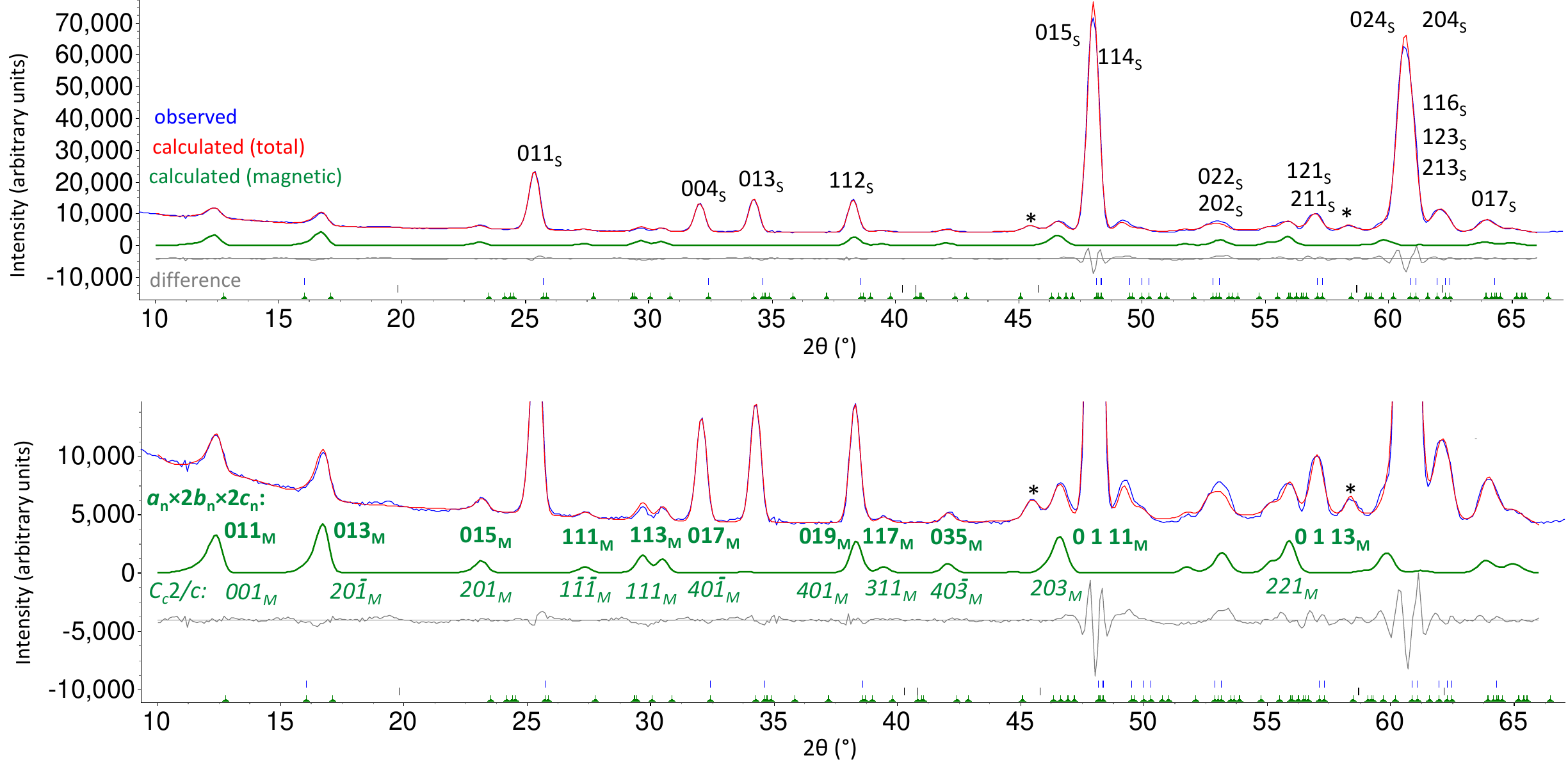}
\caption{[color online]  Rietveld refinement profiles of the 4 K data for Ce$_{2}$O$_{2}$FeSe$_{2}$ showing the observed (blue), calculated (red) and difference (grey) profiles. Both nuclear and magnetic-only phases were included in the refinement and scattering from the magnetic phase is highlighted by the solid green line. The tick marks for the nuclear structure (black, top), and the Ce$_{2}$O$_{2}$Se impurity (\textless 2\% by weight, marked by $\ast$) (blue, central) and magnetic (green, bottom) structure are shown below.   The refinement was carried out for the nuclear structure using space group $Imcb$, $a$ = 5.6788(8) \AA, $b$ = 5.7087 (9) \AA, $c$ = 17.290(2) \AA, and for the magnetic structure using space group $C_c$2/$c$, $a$ = 18.208(2) \AA, $b$ = 5.6788(8) \AA, $c$ = 11.417, $\beta$ = 108.272(3)$^{\circ}$. Moments of 3.14(8) $\mu_{B}$ and 1.14(4) $\mu_{B}$ were obtained for the Fe and Ce sites, respectively, with R$_{wp}$ = 4.34\% and R$_{p}$ = 3.23\%. $hkl$ values for nuclear reflections are given in upper panel in black; $hkl$ indices for magnetic reflections are given in lower panel in green (those for the $a_{n}$ $\times$ 2$b_{n}$ $\times$ 2$c_{n}$ magnetic unit cell above in bold; those for the $C_c$2/$c$ cell below italicised).}
\label{refinement}
\end{figure*}

Whilst the FM-chain model gives a better fit than AFM-chain models,further analysis indicated that including the Ce magnetic ordering modes improves the fit significantly ($R_{wp}$ decreases from 5.49\% to 4.37\% at 80 K, and from 7.19\% to 4.61\% at 4 K for one additional parameter). Refinements are very sensitive to the relative signs of the Fe$^{2+}$ and Ce$^{3+}$ magnetic ordering mode amplitudes. For example, as measured by $R_{wp}$, a surface plot showing fit for different amplitudes of the Ce and Fe magnetic ordering modes indicates that the best fit is obtained when both modes have the same sign, corresponding to a FM coupling between nn Fe and Ce sites (see Supplementary Material). Refinement profiles and details are shown in Fig.\ref{refinement}. If canting of the Ce moments is included in the model, the Ce moments become oriented at $\sim$12$^{\circ}$ to the $ab$ plane (i.e., a $z$ component of 0.25(5) $\mu_{B}$) and $R_{wp}$ is reduced by 0.04 \%, but this improvement cannot be regarded as significant from our data.

Analysis using ISODISTORT \cite{Campbell-2006} suggests that the magnetic structure of Ce$_{2}$O$_{2}$FeSe$_{2}$ can be described by the $C$-centered space group $C_c2/c$ [BNS: 15.9 with basis (0, -1, 1), (-1, 0, 0), (0, 2, 0) and origin at (0, 0, 0)] shown in Figure\ref{structure}$b$ and refinement using 4 K data gives moments of 3.14(8) $\mu_{B}$ and 1.14(4) $\mu_{B}$ for Fe and Ce sites, respectively. The ordered Fe$^{2+}$ moment in Ce$_{2}$O$_{2}$FeSe$_{2}$ is comparable with that reported for the Mott insulating oxyselenides (e.g., La$_{2}$O$_{2}$Fe$_{2}$OSe$_{2}$ (3.50(2) $\mu_{B}$) \cite{McCabe-2014} and the parent phase to superconducting K$_{0.8}$Fe$_{1.6}$Se$_{2}$ (3.31 $\mu_{B}$) \cite{Bao-2011}), and is consistent with a high-spin d$^{6}$ configuration for Fe$^{2+}$ sites. It is significantly larger than that observed in $Ln$FeAsO materials with poor metallic behavior (e.g. 0.94(3) $\mu_{B}$ for CeFeAsO at 1.7 K \cite{Zhao-2008}). 

The sequential Rietveld refinements using NPD data collected on cooling show that the Ce moment increases almost linearly at low temperatures. The Fe moment can be fitted well by the critical behavior with $\beta$ = 0.28(1) and $T_{\mathrm{N}}$ = 175.8(8) K (Fig.\ref{peaks}f) for 100 K $< T$ \textless  171 K. This Fe$^{2+}$ moment ordering is similar to that observed for CeFeAsO at $T_{\mathrm{N}}$ = 137 K, which can be described by critical behavior with $\beta$ = 0.24(1) \cite{Maeter-2009}. These values for $\beta$ are larger than those reported for 2D-Ising like systems (including undoped BaFe$_{2}$As$_{2}$ ($\beta$=0.125 \cite{Wilson-2009}) and La$_{2}$O$_{2}$Fe$_{2}$OSe$_{2}$ ($\beta$=0.122 \cite{McCabe-2014})), but smaller than those predicted for three-dimensional critical fluctuations ($\beta$=0.326, 0.367 and 0.345 for 3D Ising, 3D Heisenberg and 3D XY systems, respectively) \cite{Collins-1989}. The crossover between 2D and 3D universality classes has been suggested to originate from a coupling to an orbital degree of freedom \cite{Chen-2009} or the prroximity of a Lifshitz point (see, for example, Fe$_{1+x}$Te \cite{Rodriguez-2013} and BaFe$_{2}$As$_{2}$ \cite{Pajerowski-2013}).

The unusual change in the relative intensities of the different magnetic Bragg reflections observed on cooling Ce$_{2}$O$_{2}$FeSe$_{2}$ (Fig.\ref{peaks}) can be rationalised in terms of the contribution of the Ce and Fe moments to peak intensities. The magnetic modes that describe the ordering of both the Fe and Ce moments have the same basis vector $\vec{k}$ = (0 ${1\over 2}$ ${1\over 2}$). As a consequence, the ordering on these two sublattices contributes to mostly the same reflections. Based on the magnetic unit cell $a_{n}$ $\times$ 2$b_{n}$ $\times$ 2$c_{n}$, the $hkl$ reflections with $h$ = 2n, $k,l$ $\neq$ 2n and $h+k+l$ = 2n (i.e., (0 1 1), (0 1 3), (0 1 5)) have contributions from both Ce and Fe sublattices, whilst some weaker $hkl$ reflections with $h$$\neq$2n, $k,l$$\neq$2n and $h+k+l$$\neq$2n have contributions only from the Ce ordering. The ordering of the Fe and Ce sublattices adds constructively for some peak intensities (e.g., (0 1 1), (0 1 9)) and destructively for others (e.g., (0 1 5), (0 1 7), (0 1 3)) (For this latter (0 1 3) reflection the Ce contribution is small and so the intensity is dominated by Fe ordering). The non-monotonic temperature dependence of the magnetic reflection intensities observed for Ce$_{2}$O$_{2}$FeSe$_{2}$ is similar to those reported for $Ln_{2}$CuO$_{4}$ $(Ln$ = Pr, Nd, $T_{\mathrm{N, Cu}}$ = 250-325 K) \cite{Matsuda-1990} and for CeVO$_{3}$ ($T_{\mathrm{N}}$  = 124-136 K) \cite{Munoz-2003,Reehuis-2008}. 

\begin{figure*}[t] 
\includegraphics[width=0.8\linewidth,angle=0.0]{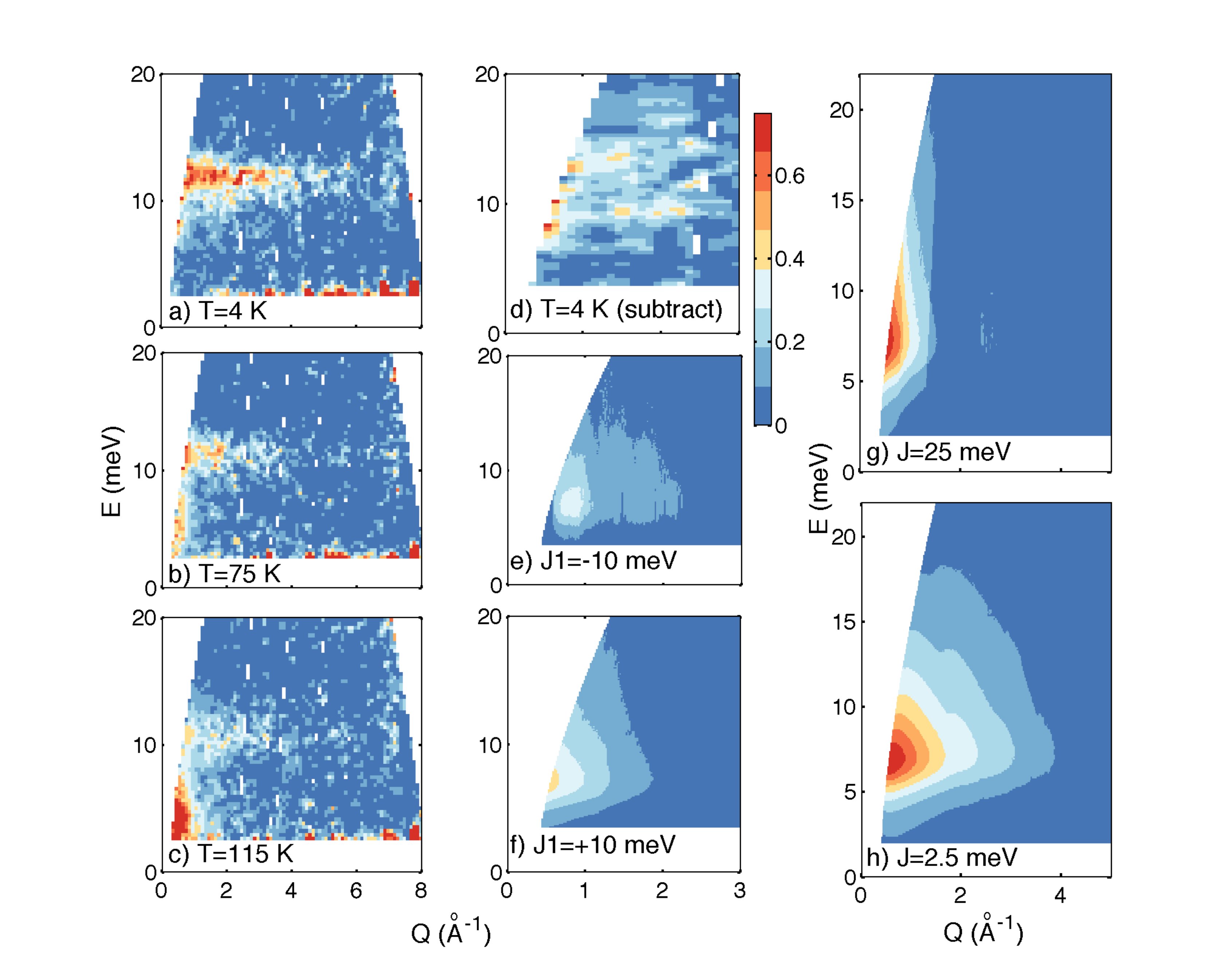}
\caption{[color online]  MARI scan with $E_{i}$ = 40 meV showing Ce CEF excitation and magnetic excitation from the Fe sublattice at (a) 4 K, (b) 75 K and (c) 115 K. (d) MARI scan with $E_{i}$ = 40 meV with scattering due to Ce CEF subtracted (see text) showing only magnetic excitation from the Fe sublattice. Powder averaged single mode analysis spin wave calculations with (e) AFM and (f) FM chains along [100]. An intrachain exchange interaction $J_{1}$ = 10 meV (positive sign denotes FM interactions) was used in these spin-wave calculations. (g), (h)  show single mode calculations for different magnitudes of the FM $J_{1}$ interaction. (The white regions at lowest momentum transfer are masked by the beam stop, and the curvature with increasing energy transfer of this inaccessible region is due to the fixed incident energy kinematics imposed by the instrument geometry.)}
\label{mari_scans}
\end{figure*}

\subsection{B. Inelastic neutron scattering}

INS was used to obtain experimental estimates for the magnetic exchange interactions.  Low-energy fluctuations were studied to probe directly the Fe-Fe exchange along the chains.  Ce crystal electric field (CEF) excitations were then investigated to determine the Fe-Ce exchange.  

Before discussing the scattering response from magnetic ions, we first describe how the background was subtracted from the powder averaged data.  The measured neutron scattering intensity $I_{meas}$ is proportional to the structure factor $S(Q, E)$ but also includes a temperature-independent background contribution due to instrument effects and sample environment. Using the principle of detailed balance, we employ data collected at different temperatures to account for the temperature-independent background. This allows us to isolate the inelastic scattering (Appendix A), which has both magnetic and lattice (phonon) contributions. To extract the purely magnetic scattering, we assume that the 300 K scattering is dominated by phonons, which is a reasonable approximation. Scaling by the Bose factor and assuming a harmonic response (Appendix 1), we estimate the phonon cross section at each temperature and subtract it from the background-corrected data. Using this method we extract the purely magnetic scattering at a given temperature, as has been used previously to study the hydrogen-containing polymeric magnet Cu(quinoxaline)Br$_{2}$ \cite{Hong-2006} and the low-energy magnetic dynamics of Fe$_{1-x}$Te$_{1-y}$Se$_{y}$ \cite{Stock-2012}. 

The purely magnetic contributions to the inelastic scattering are shown in Figure \ref{mari_scans} for $E_{i}$ = 40 meV. A strong, sharp excitation, independent of $Q$, is observed at $E$ $\sim$ 11 meV, which is ascribed to the Ce$^{3+}$ CEF excitations. At slightly lower energies, a gapped excitation is observed near $Q$ = 0. The gap value is similar to that observed in other parent Fe$^{2+}$ based superconductors such as Fe$_{1+x}$Te \cite{Stock-2011} and La$_{2}$O$_{2}$Fe$_{2}$OSe$_{2}$ \cite{McCabe-2014} and in 122 systems including BaFe$_{2}$As$_{2}$ \cite{Wilson-2010}. Unlike the crystal field excitation, this scattering is well-defined in momentum and decays quickly with momentum transfer, bearing a strong resemblance to the magnetic excitation observed in powder averaged measurements of La$_{2}$O$_{2}$Fe$_{2}$OSe$_{2}$ \cite{McCabe-2014}. The temperature-dependence of this excitation is also different from that of the Ce$^{3+}$ CEF excitation: at 4 K, it has a gap of $\sim$9 meV, which decreases on warming and softens into the elastic line by 115 K (Fig.\ref{mari_scans}$b,c$). Based on these observations, we conclude that the low energy, low-$Q$ scattering originates from the Fe$^{2+}$ magnetic sublattice.

To separate the Ce CEF excitations from the Fe$^{2+}$ magnetic excitations, the CEF contribution was estimated by taking a cut over the momentum transfer range of $Q$ = 2.8 - 3.5 \AA$^{-1}$, and then scaling by the Ce$^{3+}$ form factor \cite{Brown-2006} to estimate the momentum dependence. This subtraction takes advantage of the fact that the crystal field excitations are dispersionless and flat in momentum transfer, particularly in comparison with the strong momentum dependence of the scattering associated with the Fe sites (as observed for La$_{2}$O$_{2}$Fe$_{2}$OSe$_{2}$  \cite{McCabe-2014} for example).  This analysis leaves only the strongly momentum varying component near $Q$ = 0 (Fig.\ref{mari_scans}$d$), from which the magnetic exchange interactions between Fe$^{2+}$ sites can be estimated.  

The single-mode approximation \cite{Stock-2009} can be used to compare possible magnetic structures with different signs (AFM $J_{1}<0$, FM $J_{1}>0$) and magnitudes for the nn interaction $J_{1}$ (illustrated in Figure \ref{dft}).  Using the single mode approximation the structure factor $S(\vec{Q},E)$ can be written in terms of a momentum-dependent term $S(\vec{Q})$ and a single Dirac delta function in energy:

\begin{eqnarray}
S(\vec{Q},E)= S(\vec{Q}) \delta[E-\epsilon(\vec{Q})].
\label{zeroeth_sum}
\end{eqnarray}

\noindent where $\epsilon(\vec{Q})$ is the dispersion. We approximate $\delta(E)$ as a Lorentzian term with full-width equal to the calculated resolution width in energy. The first moment sum rule~\cite{Hohenberg-1974} relates $S(\vec{Q})$ to the dispersion:

\begin{eqnarray}
S(\vec{Q})=-{2\over 3} {1\over {\epsilon(\vec{Q})}}\sum_{\vec{d}} J_{1} \langle \vec{S}_{0} \cdot \vec{S}_{\vec{d}}\rangle [1-\cos(\vec{Q}\cdot\vec{d})].
\label{first_moment_sum}
\end{eqnarray}

\noindent where $\vec{d}$ is the bond vector connecting nn spins with an exchange interaction $J_{1}$. Making the assumption that this intrachain interaction dominates, we use the dispersion relation for the one-dimensional (1D) chain system:
\begin{eqnarray}
\epsilon(\vec{Q})^{2}=4S^{2}[\Delta^{2}+J_{1}^{2}[1-\cos(\pi H)]^{2}].
\label{disperion}
\end{eqnarray}

\noindent where $\Delta$ is the gap value determined by anisotropy and $J_{1}$ is the nn intrachain exchange interaction. 

Representative calculations using the AFM and FM chain models are summarized in Figure \ref{mari_scans}. The AFM model gives correlations at finite $Q$, whereas the FM model gives magnetic scattering only at lowest measurable wave vectors (near $Q$ = 0).  From the temperature dependence (Figure \ref{mari_scans}$a-c$) and the subtracted data (Figure \ref{mari_scans}$d$), the strongly temperature dependent magnetic scattering is present near $Q=0$. This is more consistent with a dominant FM $J_{1}$ interaction (simulated in Figure \ref{mari_scans}$f$) than an AFM interaction where the scattering is peaked at finite $Q$. Based on this comparison, we conclude that the exchange mechanism is predominately ferromagnetic ($J_{1}>0$), consistent with analysis of NPD data described above.  Figure \ref{mari_scans} $f-h$ shows results of single mode calculations for this FM chain model for various values of $J_{1}$. It is difficult to give an accurate value for this exchange interaction given the scattering is concentrated near $Q=0$, but our calculations indicate that $J_{1}$ $\sim$10-20 meV gives the best qualitative agreement with the observed data. It should be emphasized that this is an estimate of the coupling and is limited by the kinematics of the scattering geometry described above.

\begin{figure}[t] 
\includegraphics[width=1\linewidth,angle=0.0]{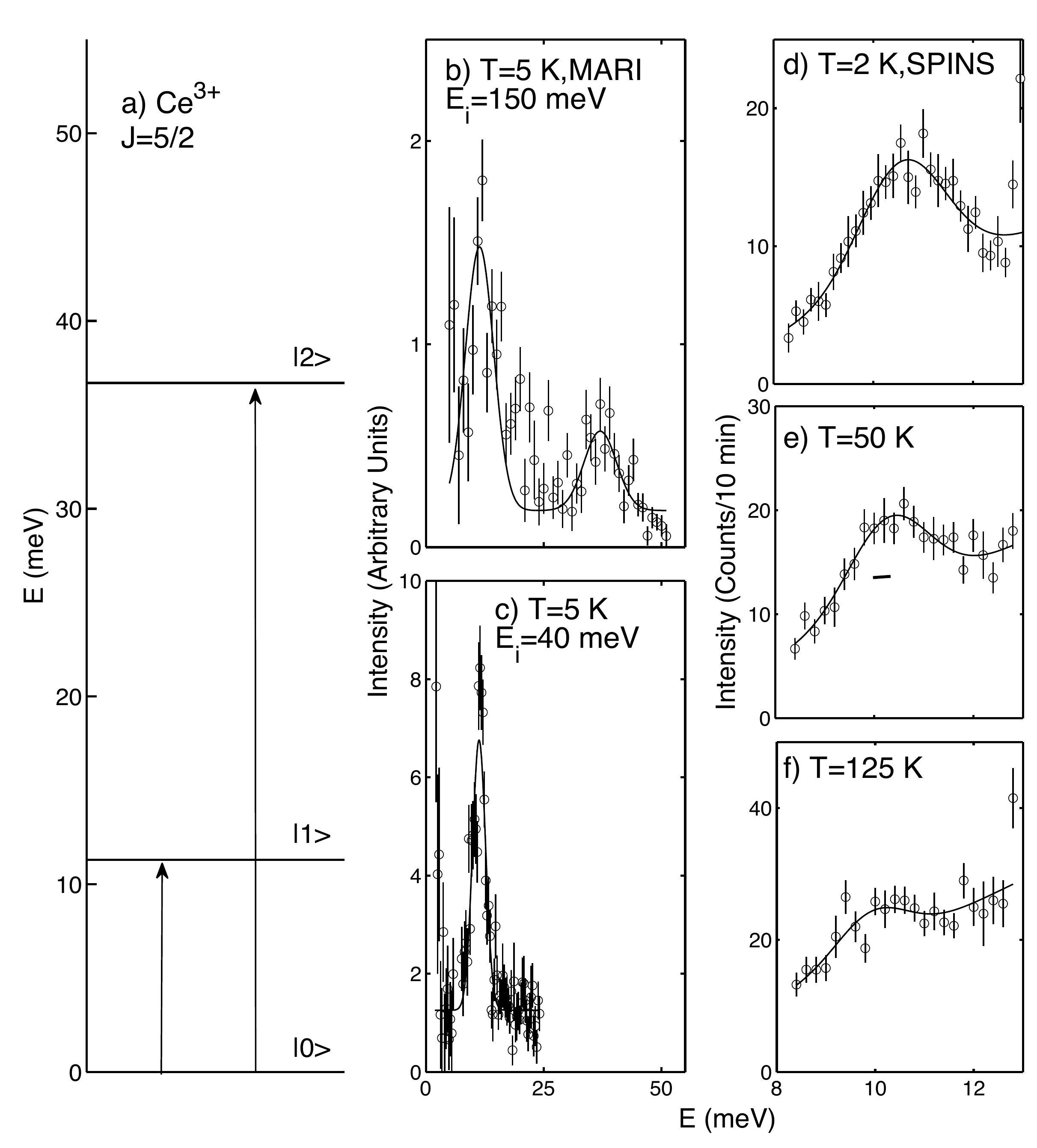}
\caption{  Mari and SPINS data for Ce$_{2}$O$_{2}$FeSe$_{2}$ showing CEF levels for Ce$^{3+}$ site as a function of temperature. The solid curves are fits to a Gaussian function.}
\label{cef}
\end{figure}

Having discussed the Fe-Fe exchange, we now discuss the localized Ce$^{3+}$ CEF excitations observed in the INS data at $\sim$11 meV and $\sim$37 meV (Fig.\ref{cef}) with the goal of extracting the coupling between Fe and Ce sites.  The magnetic nature of the peak around 11 meV is confirmed by the temperature dependence shown in Figure \ref{cef}.  The softening of the first crystal field excitation (Fig. \ref{cef}d-f) with increasing temperature could be the result of thermal expansion or of a change in the ground state ~\cite{Fulde-1978}. We note that the softening observed can be reproduced by point charge calculations and is consistent with thermal expansion. To obtain an estimate of Fe -- Ce exchange, it is important to have a heuristic model for the Ce crystal fields from which eigenfunctions and transition energies can be derived.  Ce$^{3+}$ (4f$^{1}$, $J={5\over 2}$) is a Kramers ion (Fig. \ref{cef}), and each level remains doubly degenerate for all crystalline electric fields unless a magnetic field is applied. Magnetic ordering on the iron sublattice can give rise to a molecular field at the Ce$^{3+}$ sites if there is coupling between Fe$^{2+}$ and Ce$^{3+}$ ions. In the oxyarsenide CeFeAsO, the degeneracy of Ce$^{3+}$ CEF states is lifted below $T_{N,Fe}$ suggesting some Fe - Ce coupling \cite{Chi-2008}, which is consistent with muon spin rotation spectroscopy studies \cite{Maeter-2009}. In the vacancy-ordered Ce$_{2}$O$_{2}$FeSe$_{2}$ structure (space group $Imcb$), the Ce atoms are on 8$j$ sites with local point symmetry $C_{2}$. The resulting crystal field Hamiltonian can be expressed in Stevens operators formalism which requires five nonzero terms to describe the monoclinic symmetry of the Ce$^{3+}$ site \cite{Walter-1984}:

\begin{eqnarray}
H_{C_{2}}=B_{2}^{0}O_{2}^{0} + B_{2}^{2}O_{2}^{2} + B_{4}^{0}O_{4}^{0} + B_{4}^{2}O_{4}^{2} + B_{4}^{4}O_{4}^{4}.
\label{full_hamiltonian}
\end{eqnarray}

Ideally, these five terms would be determined by fitting to the experimental data but they cannot be uniquely and unambiguously determined with only two CEF energies and intensities and the ordered moment from the NPD analysis (which depends on the Ce ground state wavefunction determined from the eigenvectors of Eqn. \ref{full_hamiltonian}).  Therefore, a simplified model for the crystal field scheme has been investigated. With no vacancy ordering on the iron sublattice (Fig.\ref{structure}), the symmetry of the Ce$^{3+}$ sites would be tetragonal and the Hamiltonan for this scheme contains only three non-zero terms:

%To guide the analysis, we used a point charge ``cluster" model (integrated over 40 unit cells to ensure convergence of the Stevens coefficients) which gave the results shown in Table \ref{table_stephens}.

%\begin{table}[ht]
%\caption{Calculated Parameters}
%\centering
%\begin{tabular} {c c}
%\hline\hline
%Stephens coefficients & value (meV)\\
%\hline\hline
%$B_{2}^{0}$ & 10.13 \\
%$B_{2}^{2}$ & 0.0039 \\
%$B_{4}^{0}$ & -0.10 \\
%$B_{4}^{2}$ & -0.0021 \\
%$B_{4}^{4}$ & -0.60 \\
%\hline
%\label{table_stephens}
%\end{tabular}
%\end{table}

\begin{eqnarray}
H_{tetrag}=B_{2}^{0}O_{2}^{0} +  B_{4}^{0}O_{4}^{0} +  B_{4}^{4}O_{4}^{4}.
\label{tetrag_hamiltonian}
\end{eqnarray}

\noindent using the two CEF energies and intensities and the ordered moment from NPD (giving five experimental ``data points"), the three coefficients are determined as $B_{2}^{0}$ = 1.5(2) meV, $B_{4}^{0}$ = -0.03(1) meV $B_{4}^{4}$ = -0.43(7) meV. The ambiguity regarding the sign of these coefficients was resolved with the results from a ``cluster" point charge calculation integrating over 40 unit cells to ensure convergence of the Stevens coefficients (see appendices for details of this calculation).

We now use this heuristic model of the crystal fields to derive an exchange coupling between the Fe -- Ce ions based upon the broadening of the crystal field levels in the magnetically ordered low temperature phase.  Because of Kramer's theorem, the crystal field excitations are doubly degenerate and only split in the presence of a time reversal violating magnetic field. This splitting can be calculated by adding the following Zeeman term to the crystal field Hamiltonian above for eigenstates $i$ and $j$:

\begin{eqnarray}
H_{Zeeman (i,j)} = \mu_{0}\mu_{B}H \langle i | J_{z} | j \rangle
\label{zeeman}
\end{eqnarray}

\noindent where $\mu_{B}$ and $\mu_{0}$ are the Bohr magneton and permeability of free space, respectively, $H$ is the effective magnetic field, $J_{z}$ is an angular momentum operator along $z$, and $\langle i | J_{z} | j \rangle$ is the angular momentum matrix element from the ground state to the excited state. To account for the powder averaging, all three directions ($x$, $y$, and $z$) were averaged. The molecular field on the Ce site is induced by magnetic ordering on the Fe sublattice (Fig.\ref{structure}$d$). In the absence of a molecular field at the Ce sites, any splitting/broadening of the Kramers doublets should arise from the Fe -- Ce coupling \cite{Chi-2008} and is consistent with muon spin relaxation studies, which indicate a strong non-Heisenberg anisotropic Fe -- Ce exchange well above $T_{\mathrm{N,Ce}}$ in CeFeAsO \cite{Maeter-2009}. In the Fe-ordered crystal structure of Ce$_{2}$O$_{2}$FeSe$_{2}$ (Fig.\ref{structure}$c$), the FM chains of edge-shared FeSe$_{4}$ tetrahedra alternate with vacant stripes along [010] and each Ce site is coupled to two Fe sites within a single FM chain (Fig.\ref{structure}$d$)) and there are no competing Fe -- Ce interactions. The molecular field on the Ce site due to the Fe magnetic sublattice is equal to 2$SJ_{4}$ where $J_{4}$ is the Fe -- Ce exchange coupling. This provides an opportunity to probe the Fe -- Ce coupling by measuring the broadening of the crystal field excitations.

The CEF levels observed for Ce$_{2}$O$_{2}$FeSe$_{2}$ are broadened (Fig.\ref{cef} $d-f$) considerably beyond the instrumental resolution (represented by the horizontal bar in Fig.\ref{cef} $e$), but it is difficult to determine the splitting of the Kramers doublets (Fig.\ref{cef}) in contrast to the case of CeAsFeO. To provide an estimate for the Fe -- Ce exchange coupling, we have fitted the low temperature excitation to a single Gaussian to obtain a full-width of 2.0(4) meV, giving a maximum value for any splitting of $\sim$1 meV. Using the Stevens parameters discussed above for the crystal field analysis, we obtain an estimate of the Fe -- Ce exchange $J_{4}$ of $\sim$0.15 meV. This estimate is approximately an order of magnitude smaller than the ferromagnetic Fe -- Fe exchange $J_{1}$ interaction discussed above.

\subsection{C. Spin exchange and electronic structure}

\begin{figure}[t] 
\includegraphics[width=1.0\linewidth,angle=0.0]{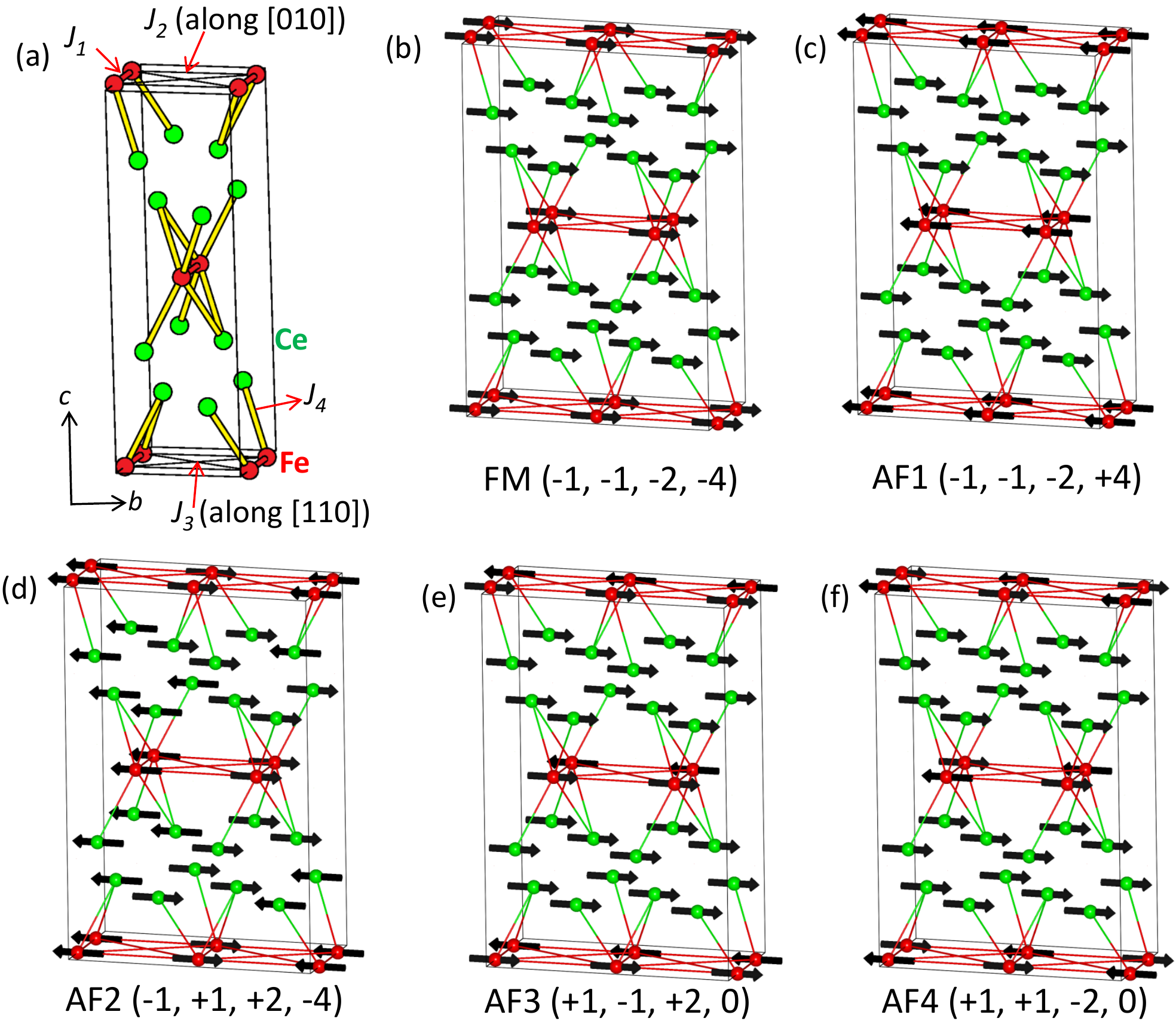}
\caption{[color online]  (a) Four spin exchange paths of Ce$_{2}$O$_{2}$FeSe$_{2}$. (b-f) Five ordered spin arrangements FM and AF1 - AF4 employed to extract $J_{1}$ -- $J_{4}$ by energy mapping analysis. The numbers in the parentheses in the first row refer to the relative energies in meV/FU, and the bracketed numbers represent the numbers $n_{1}$, $n_{2}$, $n_{3}$ and $n_{4}$ of Eq. \ref{energy}. AF2 is the structure observed experimentally from diffraction with a weak $J_{4}$ exchange expected from neutron inelastic scattering.} 
\label{dft}
\end{figure}

Summarizing the experimental results above, we observe ferromagnetic Fe -- Fe and weaker ferromagnetic Fe -- Ce exchange. This is based on both magnetic neutron diffraction and inelastic scattering results. In this section, we provide electronic structure calculations with the goal of understanding these results and comparing them with previous calculations.

Figure \ref{dft}$a$ shows the four spin exchanges of Ce$_{2}$O$_{2}$FeSe$_{2}$ we investigate, namely, the intrachain exchange,  $J_{1}$, and the interchain exchanges,  $J_{2}$ and  $J_{3}$, between Fe$^{2+}$ ions as well as the exchange  $J_{4}$ between Fe$^{2+}$ and Ce$^{3+}$ ions. To extract the values of $J_{1}$ -- $J_{4}$ by energy-mapping analysis \cite{Whangbo-2003,Dai-2005,Xiang-2013}, we consider five ordered spin states FM and AF1 -- AF4 presented in Figure \ref{dft}$b-f$. The FM, AF1 and AF2 states contain FM chains. The coupling between adjacent FM chains is FM in the FM and AF1 states, but AFM in the AF2 state. The coupling between the Fe$^{2+}$ and Ce$^{3+}$ ions is FM in the FM and AF2 states, but AFM in the AF1 state. The AF3 and AF4 states consist of AFM chains so that the net spin exchange between the Fe$^{2+}$ and Ce$^{3+}$ ions vanishes. The coupling between adjacent AFM chains is FM in the AF3 state, but AFM in the AF4 state. The AF2 state is closest to that observed experimentally. The total spin exchange energies of the FM and AF1 -- AF4 states can be expressed in terms of the spin Hamiltonian,
\begin{eqnarray}
H=-\sum_{i<j} J_{ij}\vec{S}_{i}\cdot\vec{S}_{j}.
\label{hamiltonian}
\end{eqnarray}

\noindent where $J_{ij}$ = $J_{1}$ -- $J_{4}$ is the spin exchange parameter for the interaction between the spin sites $i$ and $j$. By applying the energy expression obtained for spin dimers with N unpaired spins per spin site (four for Fe$^{2+}$, and one for Ce$^{3+}$) \cite{Dai-2001,Dai-2003}, the total spin exchange energies per formula unit (FU) of the FM and AF1 -- AF4 states can be written as
\begin{eqnarray}
E = (n_{1}J_{1} + n_{2}J_{2} + n_{3}J_{3}) \left({N_{Fe}^{2}\over 4}\right) +... \nonumber \\
 n_{4}J_{4} \left({N_{Fe}N_{Ce}\over 4}\right)
\label{energy}
\end{eqnarray}

\noindent where $N_{Fe}$ = 4 and $N_{Ce}$ = 1, and the coefficients $n_{1}$ -- $n_{4}$ for the five spin ordered states are summarized in Figure \ref{dft}$b-f$. We examined the relative energies of the FM and AF1 -- AF4 states on the basis of DFT+U electronic structure calculations with various $U_{Fe}$ and $U_{Ce}$ values. From their DFT+U calculations, Li et al \cite{Li-2012_arXiv} found that the experimentally reported magnetic structure (namely, the AF2 state) is stable for $U_{Ce}$ = 12 eV with with $U_{Fe}$ = 0. Our calculations show that structures with FM chains (AF1 and AF2) are significantly more stable than those with AFM chains (AF3 and AF4) for all combinations of $U_{Fe}$ and $U_{Ce}$ given here (Table \ref{relative_energies}) (for $U_{Fe}$ $>$ 2 eV, the nn Fe -- Se -- Fe spin exchange $J_{1}$  becomes AFM). AF1 and AF2 spin arrangements differ in the sign of Fe -- Ce exchange $J_{4}$: in AF1, Ce spins are antiparallel to nn Fe spins (i.e. AFM $J_{4}$ exchange), whereas Ce spins are parallel to nn Fe spins (i.e. FM $J_{4}$) in the AF2 arrangement. We note that the experimentally observed AF2 arrangement is the more energetically favourable for all $U_{Fe}$ and $U_{Ce}$ combinations considered, but that the relative stability of AF2 over AF1 is much greater with $U_{Ce}$ = 12 eV than with $U_{Ce}$ = 10 eV.

\begin{table}[ht]
\caption{Energies (in meV per formula unit) of AF spin arrangements shown in Figure \ref{dft} relative to FM arrangement for various $U_{Fe}$ and $U_{Ce}$ values (in eV).}
\centering
\begin{tabular} {c c c c c c}
\hline\hline
&    & AF1 & AF2 & AF3 & AF4\\
\hline\hline
$U_{Fe}$ = 2, $U_{Ce}$ = 12 &  & +1.2 & -33.0 & +149.5 & +159.8 \\
$U_{Fe}$ = 2, $U_{Ce}$ = 10 &  & -4.9 & -9.7 & +139.1 & +178.1\\
$U_{Fe}$ = 0, $U_{Ce}$ = 12 &  & +0.1 & -35.0 & +167.5 & +169.0 \\
$U_{Fe}$ = 0, $U_{Ce}$ = 10 &  & 0.0 & -5.9 & +218.1 & +222.1 \\
\hline
\label{relative_energies}
\end{tabular}
\end{table}

By mapping the relative energies of the FM and AF1 -- AF4 states, determined from the DFT+U calculations with $U_{Fe}$ = 2 eV and $U_{Ce}$ = 12 eV (see Fig. \ref{dft}$b-f$), onto the corresponding relative energies determined from Eq. \ref{energy}, we obtain $J_{1}$ = 21.3 meV, $J_{2}$ = -1.4 meV, $J_{3}$ = -1.4 meV, and $J_{4}$ = 0.2 meV (see Table \ref{table_energy_mapping}).  These spin exchanges are consistent with the observed magnetic structure of Ce$_{2}$O$_{2}$FeSe$_{2}$. The calculated value for the Fe -- Se -- Fe exchange, $J_{1}$ $\sim$ 21 meV, is comparable in magnitude to the experimental value of about 10 -- 20 meV from INS. Furthermore, the calculated value for the Fe -- Ce exchange, $J_{4}$ = 0.2 meV, is in good agreement with the experimental value of about 0.15 meV. Similar energy-mapping analyses were carried out with other values of $U_{Fe}$ and $U_{Ce}$, as shown in Table \ref{table_energy_mapping}. These calculations show that $J_{1}$ is strongly FM and that $J_{4}$ is weakly FM for the $U_{Fe}$ = 2 eV and $U_{Ce}$ = 12 eV combination (all other combinations give $J_{4}$ weakly AFM).

\begin{table}[ht]
\caption{Values of $J_{1}$ -- $J_{4}$ (in meV) from energy-mapping analyses based on various $U_{Fe}$ and $U_{Ce}$ values (in eV).}
\centering
\begin{tabular} {c c c c c c}
\hline\hline
&    & $J_{1}$ & $J_{2}$ & $J_{3}$ & $J_{4}$\\
\hline\hline
$U_{Fe}$ = 2, $U_{Ce}$ = 12 &  & +21.3 & -1.4 & -1.4 & +0.2 \\
$U_{Fe}$ = 2, $U_{Ce}$ = 10 &  & +20.7 & +1.8 & -1.5 & -0.6\\
$U_{Fe}$ = 0, $U_{Ce}$ = 12 &  & +23.2 & -2.1 & -1.1 & +0.0 \\
$U_{Fe}$ = 0, $U_{Ce}$ = 10 &  & +27.9 & -0.1 & -0.3 & 0.0 \\
\hline
\label{table_energy_mapping}
\end{tabular}
\end{table}

\begin{figure*}[t] 
\includegraphics[width=1.0\linewidth,angle=0.0]{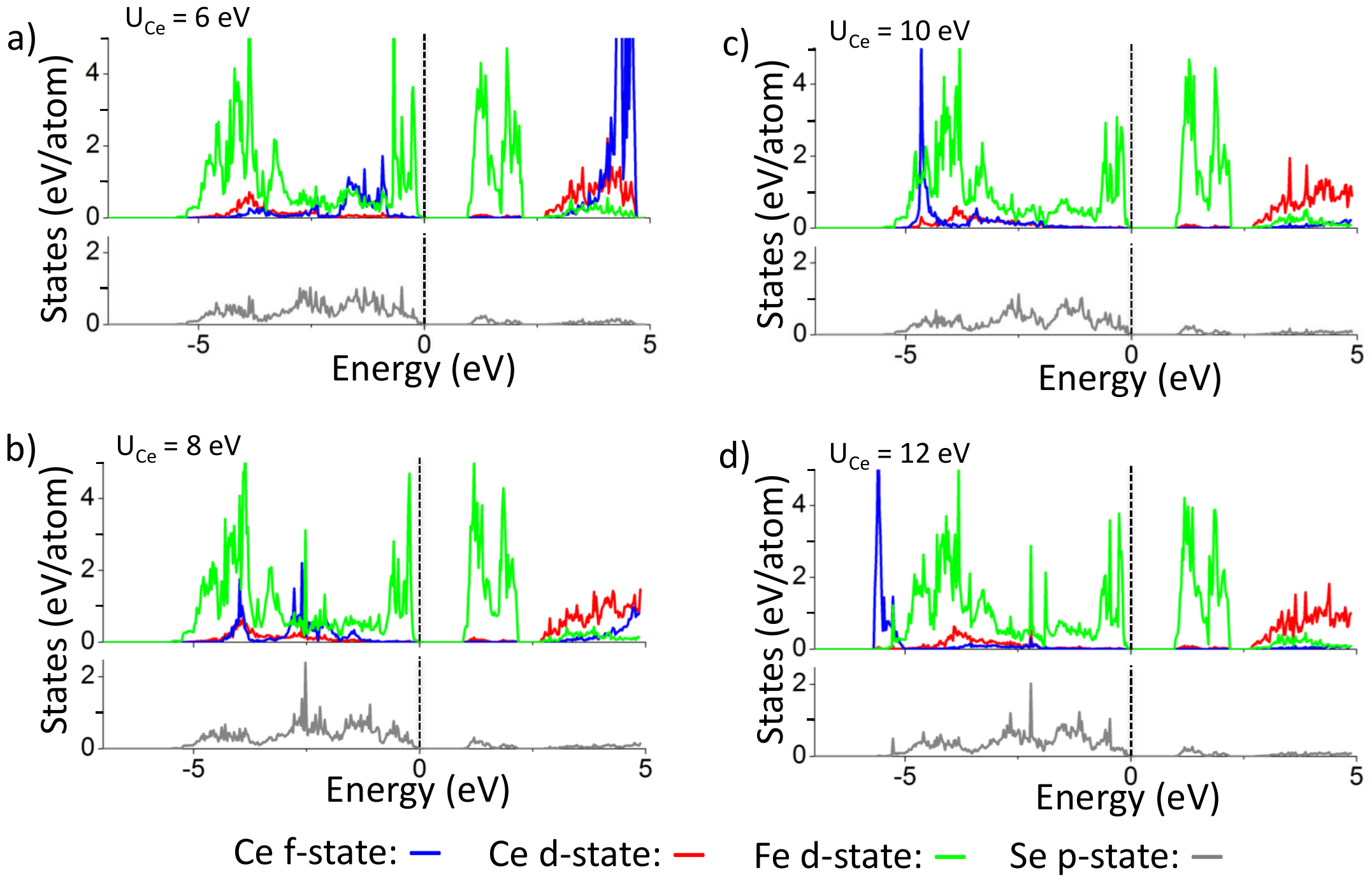}
\caption{[color online]  PDOS plots  obtained for the Ce 4f, Ce 5d, Fe 3d and Se 4p states of Ce$_{2}$O$_{2}$FeSe$_{2}$ from the DFT+U calculations with $U_{Fe}$ = 2 eV and $U_{Ce}$ = 6, 8, 10 and 12 eV. The vertical axis represents the density of states, and the horizontal axis the energy in eV.}
\label{pdos}
\end{figure*}

Figure \ref{pdos} shows plots of the projected density of states (PDOS) obtained for the Ce 4f, Ce 5d, Fe 3d and Se 4p states of Ce$_{2}$O$_{2}$FeSe$_{2}$ from the DFT+U calculations with $U_{Fe}$ = 2 eV and $U_{Ce}$ = 6, 8, 10 and 12 eV. It shows a band gap of about 1 eV which is consistent with the semiconducting behavior of Ce$_{2}$O$_{2}$FeSe$_{2}$ observed experimentally (with band gap of 0.64 eV) \cite{McCabe-2011}. 

We note from Figure \ref{pdos} that the Fe 3d states overlap with the Se 4p states througout the filled energy region, which indicates that the interaction between Fe 3d and Se 4p orbitals takes place throughout this energy range. The Ce 5d states contribute to the filled region of the Fe 3d and Se 4p states, and these contributions are not strongly affected by the change in $U_{Ce}$. However, on increasing $U_{Ce}$ from 6 eV to 12 eV, the Ce 4f states are gradually lowered in energy such that they overlap with the filled Fe 3d and Se 4p states when $U_{Ce}$ $<$ 12 eV, but do not when $U_{Ce}$ $\geq$ 12 eV. Likewise, the filled Ce 4f states overlap with the Ce 5d states when $U_{Ce}$ $<$ 12 eV but do not when $U_{Ce}$ $\geq$ 12 eV. The ferromagnetic Fe -- Se -- Ce spin exchange $J_{4}$ (and increased stability of the experimentally observed AF2 spin arrangement) is found when the Ce 5d states do not overlap in energy with the Fe 3d and Se 4p states in the energy region within 2 eV below the Fermi level. This is understandable because an antiferromagnetic Fe-Se-Ce spin exchange would involve the Fe 3d, Se 4p and Ce 5d orbitals. That the electronic structure of Ce$_{2}$O$_{2}$FeSe$_{2}$ is described by using a small value of $U_{Fe}$ (2 eV) suggests a weakly correlated nature of the iron-selenide sheets in Ce$_{2}$O$_{2}$FeSe$_{2}$. This is similar to that found for parent materials to iron-based superconductors ($U_{Fe}$ $\sim$ 2 eV for SmFeAsO and BaFe$_{2}$As$_{2}$ ~\cite{Wang-2009} and $\sim$ 4 eV used to describe the electronic properties of K$_{0.76}$Fe$_{1.72}$Se$_{2}$ ~\cite{Nekrasov-2014}. The need for a large on-site repulsion $U_{Ce}$ for Ce (12 eV) is comparable to the trend established for CeFeAsO where $U_{Ce}$ = 9 eV was required ~\cite{Pourovskii-2008}.

\section{4. Discussion}

In the low-temperature magnetic structure of Ce$_{2}$O$_{2}$FeSe$_{2}$, both Ce$^{3+}$ and Fe$^{2+}$ moments lie within the $ab$ plane, similar to the structure reported for CeFeAsO \cite{Zhao-2008}. The observation of in-plane Ce$^{3+}$ moments is consistent with the easy-axis along $x$ proposed for Ce$^{3+}$ sites in orthorhombic CeFeAsO \cite{Gornostaeva-2013}. The ordered Ce$^{3+}$ moment of Ce$_{2}$O$_{2}$FeSe$_{2}$ at 4 K (1.14(4) $\mu_{B}$) is slightly larger than that reported for CeFeAsO (0.83(2) $\mu_{B}$ at 1.7 K) \cite{Zhao-2008}, and is close to that expected for a Ce$^{3+}$ doublet ground state (1 $\mu_{B}$) \cite{Chi-2008}. The high ordering temperature for the Ce moments implied by our NPD data is surprising; other systems known to have high Ce ordering temperatures include CeRh$_{3}$B$_{2}$ (115 K) in which Ce$^{3+}$ ion is the only magnetic ion \cite{Givord-2004} and CeVO$_{3}$ (50 K) in which Ce ordering is thought to arise from FM exchange between Ce$^{3+}$ and V$^{3+}$ ions \cite{Munoz-2003,Reehuis-2008}. The high Ce$^{3+}$ moment ordering temperature in Ce$_{2}$O$_{2}$FeSe$_{2}$ is most probably due to the FM spin exchange between adjacent Ce$^{3+}$ and Fe$^{2+}$ ions, that is, the long range magnetic order of the Fe$^{2+}$ sublattice induces that of the Ce$^{3+}$ ions. In CeFeAsO, each Ce site is coupled to two FM chains of edge-sharing FeAs$_{4}$ tetrahedra with opposite spin orientations \cite{Zhao-2008}, leading to frustration of any Ce -- Fe exchange interactions \cite{Carlson-2010}. This is expected to give a negligible field on the Ce site (consistent with the low Ce moment ordering temperature). 
%The large CEF splitting reported \cite{Chi-2008} is therefore surprising and Jesche et al. report magnetic measurements on single crystals of CeFeAsO that are inconsistent with this large reported CEF splitting \cite{Jesche-2009}. 
The very small CEF splitting ($\sim$1 meV) observed for Ce$_{2}$O$_{2}$FeSe$_{2}$ is similar to that described for the parasitic ordering of Ce$^{3+}$ moments in CeMnAsO \cite{Tsukamoto-2011}. 

The nn Fe -- Fe magnetic exchange interactions, $J_{1}$, determined here experimentally are in good agreement with our DFT calculations. They are similar in magnitude to those reported for CeFeAsO \cite{Han-2009} but of opposite sign. They are significantly larger than those reported for La$_{2}$O$_{2}$Fe$_{2}$OSe$_{2}$, in which the Fe$^{2+}$ cations are coordinated by both oxide and selenide anions, which may give rise to more strongly correlated behavior \cite{McCabe-2014}. We note that the nn $J_{1}$ interactions in Ce$_{2}$O$_{2}$FeSe$_{2}$ are FM, which may reflect some orbital ordering on Fe sites, as proposed for the pnictides \cite{Chen-2009}.

\section{5. Conclusions}

In conclusion, the FM nature of Fe -- Se -- Fe nn interactions $J_{1}$ has been confirmed by NPD and INS measurements. INS work indicates that this exchange is $\sim$10 -- 20 meV. This is consistent with DFT + U calculations for $U_{Fe}$ $\le$ 2 eV and suggest that the Fe 3d electrons in the Fe$_{0.5}$Se sheets in Ce$_{2}$O$_{2}$FeSe$_{2}$ are weakly correlated, similar to the FeAs sheets in SmFeAsO and BaFe$_{2}$As$_{2}$. Weak FM Fe -- Se -- Ce interactions of 0.15 - 0.20 meV (reproduced by DFT + U calculations for $U_{Fe}$ = 2 eV) are not frustrated in this cation-ordered ZrCuSiAs-related structure. Therefore "parasitic" ordering of Ce$^{+3}$ might be induced by magnetic ordering of the Fe sublattice, with Ce$^{3+}$ moments parallel to adjacent Fe$^{2+}$ moments.

See Supplemental Material at [URL will be inserted by publisher] for details of analysis of NPD data including relative intensity of magnetic reflections on cooling as well as surface plots for magetic refinements.

We acknowledge STFC, EPSRC (EP/J011533/1), Royal Society of Edinbugh, and the NSF (DMR-0944772) for funding. We thank Emma Suard (ILL), Ross Stewart (ISIS) for assistance and Mark Green (Kent) and Efrain Rodriguez (Maryland) for helpful discussions.

\section{APPENDIX A: INS DATA ANALYSIS}

The principle of detailed balance can be used to estimate the temperature-independent background contribution to the scattering. We can approximate that for a fixed wave vector and energy transfer, the neutron energy gain (negative energy transfer, $(-|E|)$) and neutron energy loss (positive energy transfer, $(+|E|)$) are related by the following expression from the detailed balance principle:
\renewcommand{\theequation}{A1}
\begin{eqnarray}
I_{meas}(+|E|,T) = B_{1}(|E|) + S(|E|,T)
\label{A1}
\end{eqnarray}

\renewcommand{\theequation}{A2}
\begin{eqnarray}
I_{meas}(−|E|,T) = B_{2}(−|E|) + S(|E|,T)exp[{-E\over k_{B}T}]
\label{A2}
\end{eqnarray}

\noindent where $B_{1}$ and $B_{2}$ are temperature-independent background points, $S(|E|,T)$ is the scattered intensity (with both magnetic and phonon contributions) and $exp[{-E\over k_{B}T}]$ is the Boltzmann factor. We assume that the resolution of the inelastic scattering does not change over the energy range investigated. With data collected at two or more temperatures, $B_{1}$ and $B_{2}$ can be determined. For $E_{i}$ = 40 meV, data were collected at six temperatures (4 K, 75 K, 115 K, 150 K, 200 K and 300 K). These data give us experimental data points in both the energy gain and energy loss spectra (giving 12 data points in total) with which the two background points $B_{1}$ and $B_{2}$ and the six values for $S(|E|,Q,T)$ can be determined.

This detailed balance allows us to isolate the inelastic scattering but this has contributions from both magnetic and phonon scattering. The measured intensity $I_{meas}$ is proportional to the structure factor $S(Q, E)$ which is related to the imaginary part of the susceptibility $\chi''(Q,E)$:
\renewcommand{\theequation}{A3}
\begin{eqnarray}
I_{meas}\propto S(Q,E) = {1\over \pi } [n(E)+1] \chi''(Q,E)
\label{A3}
\end{eqnarray}

\noindent where n(E) is the Bose factor. The scattering at 300 K is dominated by phonons and so the phonon contribution $\chi''_{phonon}(Q,E)$ can be written:
\renewcommand{\theequation}{A4}
\begin{eqnarray}
\chi''_{phonon}(Q,E) = {S_{300 K}(Q,E)\over [n(E)_{300 K}+1] }
\label{A4}
\end{eqnarray}

The phonon contribution at each temperature was then estimated using equation A3 and subtracted to obtain the purely magnetic scattering at each temperature.

\section{APPENDIX B: POINT CHARGE CLUSTER MODEL CALCULATION}
To guide the CEF analysis, we used a point charge ``cluster" model (integrated over 40 unit cells to ensure convergence of the Stevens coefficients) which gave the results shown in Table \ref{table_stephens}.

\renewcommand{\thetable}{B1}
\begin{table}[ht]
\caption{Calculated Parameters}
\centering
\begin{tabular} {c c}
\hline\hline
Stevens coefficients & value (meV)\\
\hline\hline
$B_{2}^{0}$ & 10.13 \\
$B_{2}^{2}$ & 0.0039 \\
$B_{4}^{0}$ & -0.10 \\
$B_{4}^{2}$ & -0.0021 \\
$B_{4}^{4}$ & -0.60 \\
\hline
\label{table_stephens}
\end{tabular}
\end{table}


\begin{thebibliography}{10}
\bibitem{Kamihara-2008} Y.~Kamihara, T.~Watanabe, M.~Hirano and H.~Hosono, J. Am. Chem. Soc. \textbf{130}, 3296 (2008).
\bibitem{Paglione-2010} J. Paglione and R. L. Greene, Nat. Phys. \textbf{6}, 645 (2010).
\bibitem{Stewart-2011} G. R. Stewart, Rev. Mod. Phys. \textbf{83}, 1859 (2011).
\bibitem{Johnston-2010} D. C. Johnston, Adv. Phys. \textbf{59}, 803 (2010).
\bibitem{Ren-2008} Z.-A.~Ren, W.~Wu, J.~Yang, W.~Yi, X.-L.~Shen, Z.-C.~Li, G.-C.~Che, X.-L.~Dong, L.-L.~Sun, F.~Zhou, Z.-X.~Zhao, Chin. Phys. Lett. \textbf{25}, 2215 (2008).
\bibitem{Wysocki-2011} A.~L.~Wysocki, D.~D.~Belashchenko, V.~P.~Antipov, Nat. Phys. \textbf{7}, 485 (2011).
\bibitem{Birgeneau-2006} R. J. Birgeneau, C. Stock, J. M. Tranquada, K. Yamada, J. Phys. Soc. Jpn. \textbf{11}, 111003 (2006). 
\bibitem{Kastner-1998} M. A. Kastner, R. J. Birgeneau, G. Shirane, Y. Endoh, Rev. Mod. Phys. \textbf{70}, 897 (1998).
\bibitem{Johnson-1974} V.~Johnson, W.~Jeitschko, J. Solid State Chem. \textbf{11}, 161 (1974).
\bibitem{Ban-1965} Z.~Ban, M.~Sikirica, Acta Cryst. \textbf{18}, 594 (1965).
\bibitem{Cruz-2008} C.~de la Cruz, Q.~Huang, J.~W.~Lynn, J.~Li, W.~Ratcliff, K.~L.~Zaretsky, H.~A.~Mook, G.~F.~Chen, J.~L.~Luo, N.~L.~Wang, P.~Dai, Nature \textbf{453}, 899 (2008).
\bibitem{Goldman-2008} A.~I.~Goldman, D.~N.~Argyriou, B.~Ouladdiaf, T.~Chatterji, A.~Kreyssig, S.~Nandi, N.~Ni, S.~L.~Bud'ko, P.~C.~Canfield, R.~J.~McQueeney, Phys. Rev. B \textbf{78}, 100506 (2008).
\bibitem{Rodriguez-2010} E.~E.~Rodriguez, P.~Zavalij, P.-Y. Hsieh, M.~A.~Green, J. Am. Chem. Soc. \textbf{132}, 10006 (2010).
\bibitem{Mizuguchi-2009} Y.~Mizuguchi, F.~Tomioka, S.~Tsuda, T.~Yamaguchi, Y.~Takano, Appl. Phys. Lett. \textbf{94}, 012503 (2009).
\bibitem{Hsu-2008} F.-C.~Hsu, J.-Y.~Luo, K.-W.~Yeh, T.-K.~Cen, T.~W.~Huang, P.~M.~Wu, Y.-C.~Lee, Y.-L.~Huang, Y.~Y.~Chu, D.-C.~Yan, M.-K.~Wu, PNAS \textbf{105}, 14262 (2008).
\bibitem{Margadonna-2009} S.~Margadonna, Y.~Takabayashi, Y.~Ohishi, Y.~Mizuguchi, Y.~Takano, T.~Kagayama, T.~Nakagawa, M.~Takata, K.~Prassides, Phys. Rev. B \textbf{80}, 064506 (2009).
\bibitem{Guo-2010} J.~Guo, S.~Jin, G.~Wang, S.~Wang, K.~Zhu, T.~Zhou, M.~He, X.~Chen, Phys. Rev. B \textbf{82}, 180520(R) (2010).
\bibitem{Yan-2011} X.-W.~Yan, M.~Gao, Z.-Y.~Lu, T.~Xiang, Phys. Rev. B \textbf{83}, 233205 (2011).
\bibitem{Lei-2011} H.~Lei, M.~Abeykoon, E.~S.~Bozin, C.~Petrovic, Phys. Rev. B \textbf{83}, 180503(R) (2011).
\bibitem{Fang-2012} M. -H. Fang, H.-D. Wang, C.-H. Dong, Z.-J. Li, C.-M. Feng, J. Chen, and H. Q. Yuan, Europhys. Lett. \textbf{94}, 27009 (2011).
\bibitem{Zhao-2012} J.~Zhao, H.~Cao, E.~Bourret-Courchesne, D.-H.~Lee, R.~J.~Birgeneau, Phys. Rev. Lett. \textbf{109}, 267003 (2012).
\bibitem{Petrovic-2011} N.~Lazarevic, H.~Lei, C.~Petrovic, Z.~V.~Popovic,  Phys. Rev. B \textbf{84}, 214305 (2011).
\bibitem{Petrovic-2011b} K.~Wang, H.~Lei, C.~Petrovic,  Phys. Rev. B \textbf{83}, 174503 (2011).
\bibitem{Bao-2011} W.~Bao, Q.-Z.~Huang, G.-F.~Chen, M.~A.~Green, D.-M.~Wang, J.-B.~He, Y.-M.~Qiu, Chin. Phys. Lett. \textbf{28}, 086104 (2011).
\bibitem{Zhao-2008} J.~Zhao, Q.~Huang, C.~de la Cruz, S.~Li, J.~W.~Lynn, Y.~Chen, M.~A.~Green, G.~F.~Chen, G.~Li, Z.~Li, J.~L.~Luo, N.~L.~Wang, P.~Dai, Nat. Mat. \textbf{7}, 953 (2008).
\bibitem{Fang-2008} C.~Fang, H.~Yao, W.-F.~Tsai, J.~P~Hu, S.~A.~Kivelson, Phys. Rev. B \textbf{77}, 224509 (2008).
\bibitem{Yildirim-2008} T.~Yildirim, Phys. Rev. Lett. \textbf{101}, 057010 (2008).
\bibitem{Si-2008} Q.~Si, E.~Abrahams, Phys. Rev. Lett. \textbf{101}, 076401 (2008).
\bibitem{Li-2012} T.~J.~Li, Y.-M.~Quan, D.-Y.~Liu, L.-J.~Zou, J. Magn. Magn. Mater. \textbf{324}, 1046 (2012).
\bibitem{Maeter-2009} H.~Maeter, H.~Luetkens, Y.~G.~Pashkevich, A.~Kwadrin, R.~Khasanov, A.~Amato, A.~A.~Gusev, K.~V.~Lamonova, D.~A.~Chervinskii, R.~Klingeler, C.~Hess, G.~Behr, B.~Buchner, H.-H.~Klauss, Phys. Rev. B \textbf{80}, 094524 (2009).
\bibitem{Zhang-2013} Q.~Zhang, W.~Tian, H.~Li, J.-W.~Kim, J.~Yan, R.~W.~McCallum, T.~A.~Lograsso, J.~L.~Zarestky, S.~L.~Bud'ko, R.~J.~McQueeney, D.~Vaknin, Phys. Rev. B \textbf{88}, 174517 (2013).
\bibitem{Bruning-2008} E.~M.~Bruning, C.~Krellner, M.~Baenitz, A.~Jesche, F.~Steglich, C.~Geibel, Phys. Rev. Lett. \textbf{101}, 117206 (2008).
\bibitem{Krellner-2007} C.~Krellner, N.~S.~Kini, E.~M.~Bruning, K.~Koch, H.~Rosner, M.~Nicklas, M.~Baenitz, C.~Geibel, Phys. Rev. B \textbf{76}, 104418 (2007).
\bibitem{McCabe-2011} E.~E.~McCabe, D.~G.~Free, J.~S.~O.~Evans, Chem. Commun. \textbf{47}, 1261 (2011).
\bibitem{Goodenough-1963} J.~B.~Goodenough, \textit{Magnetism and the chemical bond}, John Wiley and Sons, New York-London, (1963).
\bibitem{Goodenough-1955} J.~B.~Goodenough, Phys. Rev. \textbf{100} 564 (1955).
\bibitem{Kanamori-1959} J.~Kanamori, J. Phys. Chem. Solids \textbf{10} 87 (1959).
\bibitem{Rietveld-1969} H.~M.~Rietveld, J. Appl. Cryst. \textbf{2}, 65 (1969).
\bibitem{Coelho-2003} A.~A.~Coelho, J. Appl. Cryst. \textbf{36}, 86 (2003).
\bibitem{Coelho-2012} A.~A.~Coelho, \textit{TopasAcademic: general profile and structure analysis software for powder diffraction data}, Bruker AXS: Karlsruhe, Germany (2012).
\bibitem{Campbell-2006} B.~J.~Campbell, H.~T.~Stokes, D.~E.~Tanner, D.~M.~Hatch, J. Appl. Cryst. \textbf{39}, 607 (2006).
\bibitem{Kresse-1993} G.~Kresse, J.~Hafner, Phys. Rev. B \textbf{47}, 558 (1993).
\bibitem{Kresse-1996} G.~Kresse, J.~Furthm{\"u}ller, Computational Materials Science \textbf{6} 15 (1996).
\bibitem{Kresse-1996b} G.~Kresse, J.~Furthm{\"u}ller, Phys. Rev. B \textbf{54}, 11169 (1996).
\bibitem{Perdew-1996} J.~P.~Perdew, K.~Burke, M.~Ernzerhof, Phys. Rev. Lett. \textbf{77}, 3865 (1996).
\bibitem{Dudarev-1998} S.~L.~Dudarev, G.~A.~Botton, S.~Y.~Savrasov, C.~J.~Humphreys, A.~P.~Sutton, Phys. Rev. B \textbf{57}, 1505 (1998).
\bibitem{Tuxworth-2013} A.~J.~Tuxworth, E.~E.~McCabe, D.~G.~Free, S.~J.~Clark, J.~S.~O.~Evans, Inorg. Chem. \textbf{52}, 2078 (2013).
\bibitem{Bhoi-2011} D.~Bhoi, P.~Mandal, P.~Choudhury, S.~Pandya, V.~Ganesan, J. Appl. Phys. \textbf{110}, 113722 (2011).
\bibitem{McCabe-2014} E.~E.~McCabe, C.~Stock, E.~E.~Rodriguez, A.~S.~Wills, J.~W.~Taylor, J.~S.~O.~Evans, Phys. Rev. B \textbf{89}, 100402(R) (2014).
\bibitem{Wilson-2009} S.~D.~Wilson, Z.~Yamani, C.~R.~Rotundu, B.~Freelon, E.~Bourret-Courchesne, R.~J.~Birgeneau, Phys. Rev. B \textbf{79}, 184519 (2009).
\bibitem{Collins-1989} M.~Collins, \textit{Magnetic critical scattering (Oxford series on neutron scattering in condensed matter)}, Oxford University Press, New York (1989).
\bibitem{Chen-2009} C.-C.~Chen, B.~Moritz, J.~van~den~Brink, T.~P.~Devereaux, R.~R.~P.~Singh, Phys. Rev. B \textbf{80}, 180418(R) (2009).
\bibitem{Rodriguez-2013} E.~E.~Rodriguez, D.~A.~Sokolov, C.~Stock, M.~A.~Green, O.~Sobolev, J.~A.~Rodriguez-Rivera, H.~Cao, A.~Daoud-Aladine, Phys. Rev. B \textbf{88}, 165110 (2013).
\bibitem{Pajerowski-2013} D.~M.~Pajerowski, C.~R.~Rotundu, J.~W.~Lynn, R.~J.~Birgeneau, Phys. Rev. B \textbf{87}, 134507 (2013).
\bibitem{Matsuda-1990} M.~Matsuda, K.~Yamada, K.~Kakurai, H.~Kadowaki, T.~R.~Thurston, Y.~Endoh, Y.~Hidaka, R.~J.~Birgeneau, M.~A.~Kastner, P.~M.~Gehring, A.~H.~Moudden, G.~Shirane, Phys. Rev. B \textbf{42}, 10098 (1990).
\bibitem{Munoz-2003} A.~Munoz, J.~A.~Alonso, M.~T.~Casais, M.~J.~Martinez-Lope, J.~L.~Martinez, M.~T.~Fernandez-Diaz, Phys. Rev. B \textbf{68}, 144429 (2003).
\bibitem{Reehuis-2008} M.~Reehuis, C.~Ulrich, P.~Pattison, M.~Miyasaka, Y.~Tokura, B.~Keimer, Eur. Phys. J. B \textbf{64}, 27 (2008).
\bibitem{Hong-2006} T.~Hong, M.~Kenzelmann, M.~M.~Turnbull, C.~P.~Landee, B.~D.~Lewis, K.~P.~Schmidt, G.~S.~Uhrig, Y.~Qiu, C.~Broholm, D.~Reich, Phys. Rev. B \textbf{74}, 094434 (2006).
\bibitem{Stock-2012} C.~Stock, E.~E.~Rodriguez, M.~A.~Green, Phys. Rev. B \textbf{85}, 094507 (2012).
\bibitem{Stock-2011} C.~Stock, E.~E.~Rodriguez, M.~A.~Green, P.~Zavalij, J.~A.~Rodriguez-Rivera, Phys. Rev. B \textbf{84}, 045124 (2011).
\bibitem{Wilson-2010} S.~D.~Wilson, Z.~Yamani, C.~R.~Rotundu, B.~Freelon, P.~N.~Valdivia, E.~Bourret-Courchesne, J.~W.~Lynn, S.~Chi, T.~Hong, R.~J.~Birgeneau, Phys. Rev. B \textbf{82}, 144502 (2010).
\bibitem{Brown-2006} P.~J.~Brown, \textit{Magnetic Form Factors} in \textit{International Tables for Crystallography} \textbf{C}, edited by A.~J.~C.~Wilson, Kluwer Academic Publishers: Dordrecht, The Netherlands (2006).
\bibitem{Stock-2009} C.~Stock, L.~C.~Chapon, O.~Adamopoulos, A.~Lappas, M.~Giot, J.~W.~Taylor, M.~A.~Green, C.~M.~Brown, P.~G.~Radaelli, Phys. Rev. Lett. \textbf{103} 077202 (2009).
\bibitem{Hohenberg-1974} P.~C.~Hohenberg, W.~F.~Brinkman, Phys. Rev. B \textbf{10}, 128 (1974).
\bibitem{Fulde-1978} P.~Fulde, \textit{Crystal Fields} in \textit{Handbook on the Physics and Chemistry of Rare Earths}, edited by K.~A.~Gschneidner and L.~Eyring,  North-Holland Publishing Company, p.279 (1978).
\bibitem{Chi-2008} S.~Chi, D.~T.~Adroja, T.~Guidi, R.~Bewley, S.~Li, J.~Zhao, J.~W.~Lynn, C.~M.~Brown, Y.~Qiu, G.~F.~Chen, J.~L.~Lou, N.~L.~Wang, P.~Dai, Phys. Rev. Lett. \textbf{101}, 217002 (2008).
\bibitem{Walter-1984} U.~Walter, J. Phys. Chem. Solids \textbf{45} 401 (1984).
\bibitem{Whangbo-2003} M.-H.~Whangbo, H.-J.~Koo, D.~Dai, J. Solid State Chem. \textbf{176} 417 (2003).
\bibitem{Dai-2005} D.~Dai, M.-H.~Whangbo, H.-J.~Koo, X.~Rocquefelte, S.~Jobic, A.~Villesuzanne, Inorg. Chem. \textbf{44} 2407 (2005).
\bibitem{Xiang-2013} H.~J.~Xiang, C.~Lee, H.-J.~Koo, X.~G.~Gong, M.-H.~Whangbo, Dalton Trans. \textbf{42}, 823 (2013).
\bibitem{Dai-2001} D.~Dai, M.-H.~Whangbo, J. Chem. Phys. \textbf{114} 2887 (2001).
\bibitem{Dai-2003} D.~Dai, M.-H.~Whangbo, J. Chem. Phys. \textbf{118} 29 (2003).
\bibitem{Li-2012_arXiv} W.~Li, C.~Setty, X.~H.~Chen, J.~Hu, Front. Phys. DOI 10.1007/s11467-014-0428-y (2014).
\bibitem{Wang-2009} W.~L.~Yang, A.~P.~Sorini, C.-C.~Chen, B.~Moritz, W.~S.~Lee, F.~Vernay, P.~Olalde-Velasco, J.~D.~Denlinger, B.~Delley, J.-H.~Chu, J.~G.~Analytis, I.~R.~Fisher, Z.~A.~Ren, J.~Yang, W.~Lu, Z.~X.~Zhao, J.~van~den~Brink, Z.~Hussain, Z.-X.~Shen, T.~P.~Devereaux, Phys. Rev. B \textbf{80}, 014508 (2009).
\bibitem{Nekrasov-2014} I.~A.~Nekrasov, M.~V.~Sadovskii, J. Exp. Theor. Phys. \textbf{99}, 598 (2014).
\bibitem{Pourovskii-2008} L.~Pourovskii, V.~Vildosola, S.~Biermann, A.~Georges, Eur. Phys. Lett. \textbf{84} 37006 (2008).
\bibitem{Gornostaeva-2013} O.~V.~Gornostaeva, K.~V.~Lamonova, S.~M.~Orel, Y.~G.~Pashkevich, Low Temp. Phys. \textbf{39} 343 (2013).
\bibitem{Givord-2004} F.~Givord, J.-X.~Boucherie, E.~Lelievre-Berna, P.~Lejay, J. Phys. Condens. Matter \textbf{16} 1211 (2004).
\bibitem{Carlson-2010} S.-L.~Li, D.-X.~Yao, Y.-M.~Qiu, H.~J.~Kang, E.~W.~Carlson, J.-P.~Hu, G.-F.~Chen, N.-L.~Wang, P.-C.~Dai, Front. Phys. China \textbf{5} 161 (2010).
\bibitem{Tsukamoto-2011} Y.~Tsukomoto, Y.~Okamoto, K.~Matsuhira, M.-H.~Whangbo, Z.~Hiroi, J. Phys. Soc. Jpn. \textbf{80} 094708 (2011).
\bibitem{Han-2009} M.~J.~Han, Q.~Yin, W.~E.~Pickett, S.~Y.~Savrasov, Phys. Rev. Lett. \textbf{102}, 107003 (2009).

\end{thebibliography}
\end{document}